\documentclass[conference]{IEEEtran}
\IEEEoverridecommandlockouts
\usepackage[utf8]{inputenc}
\title{A multi-label, dual-output deep neural network for automated bug triaging}
 \author{Christopher A. Choquette-Choo,\textsuperscript{1,5}
David Sheldon,\textsuperscript{2}
Jonny Proppe,\textsuperscript{3,4}
John Alphonso-Gibbs,\textsuperscript{2}
Harsha Gupta\textsuperscript{2}\\
\textsuperscript{1}{Work completed while at Intel PSG, San Jose, California, United States}\\
\textsuperscript{2}{Intel PSG, San Jose, California, United States}\\
\textsuperscript{3}{University of Toronto, Departments of Chemistry and Computer Science, Toronto, Ontario, Canada}\\
\textsuperscript{4}{\textit{Current address}: University of Göttingen, Institute of Physical Chemistry, Göttingen, Germany}}

\usepackage{fancyhdr}
\fancypagestyle{empty}{%
  \fancyhf{}
  \fancyfoot[C]{\thepage}
  
}

\usepackage[numbers,sort]{natbib}
\usepackage[hang,flushmargin]{footmisc}
\usepackage{amsmath}

\usepackage{etoolbox}
\makeatletter
\patchcmd{\@makecaption}
  {\scshape}
  {}
  {}
  {}
\makeatother
\usepackage{algorithm}
\usepackage{algpseudocode}
\usepackage{amssymb}
\usepackage{mathtools}
\usepackage{tabularx}
\makeatletter
\def\footnoterule{\kern-3\p@
  \hrule \@width 2in \kern 2.6\p@} % the \hrule is .4pt high
\makeatother
\makeatletter
\usepackage{times}
\usepackage{helvet}
\usepackage{courier}
\usepackage[hyphens]{url}
\usepackage{graphicx}
\newcommand{\pml}{\hspace{0.75mm}\pm<\hspace{-0.75mm}}

\begin{document}

\maketitle

\thispagestyle{plain}
\pagestyle{plain}
\section{Abstract}
Bug tracking enables the monitoring and resolution of issues and bugs within organizations. Bug triaging, or assigning bugs to the owner(s) who will resolve them, is a critical component of this process because there are many incorrect assignments that waste developer time and reduce bug resolution throughput. In this work, we explore the use of a novel two-output deep neural network architecture (Dual DNN) for triaging a bug to both an individual team and developer, simultaneously. Dual DNN leverages this simultaneous prediction by exploiting its own guess of the team classes to aid in developer assignment. A multi-label classification approach is used for each of the two outputs to learn from all interim owners, not just the last one who closed the bug. We make use of a heuristic combination of the interim owners (owner-importance-weighted labeling) which is converted into a probability mass function (pmf). We employ a two-stage learning scheme, whereby the team portion of the model is trained first and then held static to train the team--developer and bug--developer relationships. The scheme employed to encode the team--developer relationships is based on an organizational chart (org chart), which renders the model robust to organizational changes as it can adapt to role changes within an organization. There is an observed average lift (with respect to both team and developer assignment) of 13\%-points in 11-fold incremental-learning cross-validation (IL-CV) accuracy for Dual DNN utilizing owner-weighted labels compared with the traditional multi-class classification approach. Furthermore, Dual DNN with owner-weighted labels achieves average 11-fold IL-CV accuracies of 76\% (team assignment) and 55\% (developer assignment), outperforming reference models by 14\%- and 25\%-points, respectively, on a proprietary dataset with 236,865 entries.

% Footnotes 
\addtocounter{footnote}{5}\footnotetext{Corresponding author: christopher.choquette.choo@mail.utoronto.ca}
% {\footnote[5]{}}
{\let\thefootnote\relax\footnote{© 2019 IEEE. Personal use of this material is permitted.
Permission from IEEE must be
obtained for all other uses,
in any current or future media, including
reprinting/republishing this material for advertising or promotional purposes, creating new
collective works, for resale or redistribution to servers or lists, or reuse of any copyrighted
component of this work in other works.}}

\section{Introduction}
Bug tracking is a vital and time-sensitive process in many organizations because it enables easy consolidation and distribution of different requests. Bug assignment, or triaging, is a critical step in this process, as the difficulties in determining the most suitable developer can become a bottleneck \cite{information_extraction_methods,reducing_effort_triage,lsiapplied} (see Section~\ref{problem}). These inefficiencies stem from the triager's imperfect predictive abilities for a developer's technical specialty, which may be further hindered by intra-organizational role changes \cite{information_extraction_methods,reducing_effort_triage,improve_bug_triage}. Therefore, determining the most suitable developer to own a bug may require several attempts; each one wastes developer time to diagnose a bug not in their domain of specialty and overall, decreases bug resolution throughput. These inefficiencies have sparked research in automated approaches supported by supervised machine learning to accurately assigning bugs to suitable developers:

Naive Bayes and Bayesian networks achieved top-1 accuracies of up to 32\% when combined with bug tossing \cite{nb_svm_bug_tossing,who_should_fix_bug}; stacked generalization ensembles were shown to increase accuracy by 1\%- to 8\%-points compared with the best individual classifier \cite{ensemble}. Researchers have also explored topic modeling approaches which show further improved accuracy \cite{tailored_1,tailored_2}. More recently, a deep, bi-directional recurrent neural network showed improvements over bag-of-words (BOW) \cite{deep_triage}. These approaches explore triaging of human-created bug reports whereas this paper will focus on automatically generated bugs from failing regression tests, which are common corporate software tests. Since these bugs are automatically generated, the associated failure data (see Section~\ref{data}) is used instead of human-typed bug titles and descriptions.

Previous studies employed both multi-class classification approaches \cite{improve_bug_triage, deep_triage, who_should_fix_bug, nb_svm_bug_tossing, ensemble, tossinggraphs, developerprioritization, semisupervised} with a single label (often the final owner, also termed closer) and (relatively fewer) multi-label approaches taking into account either the set \cite{multilabelknn} or the sequence \cite{seqtriage} of developers associated with a bug.
Choosing the closer as the label in multi-class approaches is built on the assumption that they are the most suitable owner for a given bug, which was raised as a potential issue by Mani et al.\ \cite{deep_triage}. Multi-label approaches can resolve this issue. However, in the examples mentioned \cite{multilabelknn, seqtriage}, the actual contributions made by specific developers were not taken into consideration. These contributions are important because each developer has varying levels of impact on a case --- all of which is implicitly contained in a bug's history (see Section~\ref{problem}).
Furthermore, some multi-class classification studies have explored using an external bug-tossing graph to enable a quasi-multi-label approach \cite{tossinggraphs,improve_bug_triage,nb_svm_bug_tossing,developerprioritization}. Each node of the graph represents a developer and every directed edge is associated with the probability of a developer tossing the bug to another developer. The graph is utilized after a multi-class classifier to enable a quasi-multi-label approach, which improved the baseline accuracy by up to 12\%-points in the examples mentioned. Extensions of this work include considering bug-specific information and the number of developer references \cite{tossinggraphs,developerprioritization}.

In this paper, we explore the synergy between
\begin{itemize}
    \item a novel two-output DNN architecture (see Section~\ref{DNN}), which exploits its own knowledge of the team classes to improve on its developer predictions, and
    \item a multi-label classification approach using a heuristic to learn from both a bug's owner history and each developer's level of contribution (see Section~\ref{heuristic}).
\end{itemize}

Finally, to render the model robust toward organizational changes, we encoded the output layers in tandem with a graphical representation of the org chart (see Section~\ref{OrgChart}).

\section{Problem Motivation}\label{problem}
\begin{figure}[h!]
\centering
\includegraphics[scale=0.57]{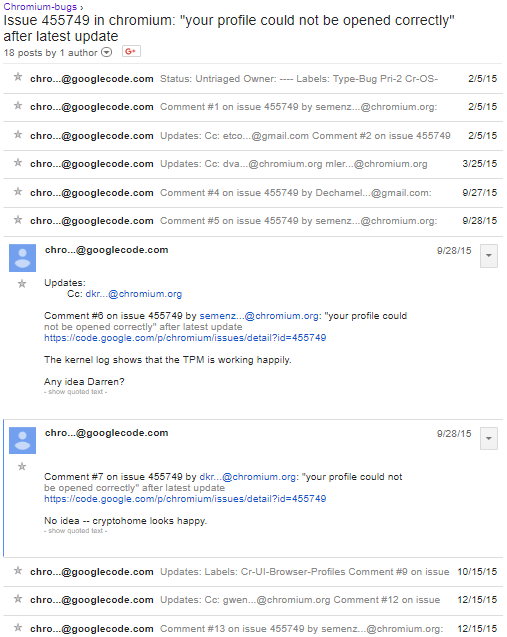} 
\caption{Screenshot of bug 455,749 on the Google Chromium project that took over 2 years and more than 30 developers to be resolved.
}
\label{fig:chromium}
\end{figure}

In bug 455,749 on the Google Chromium project (Figure~\ref{fig:chromium}), an internal developer encountered an issue with logging in. The developer tried assigning their bug to a fellow developer that they believed could fix the problem. As shown in the expanded portions of Figure~\ref{fig:chromium}, this developer did not know what the cause or solution was. This process repeated many times with over 30 developers touching the bug, each one took an average of 38 days to respond and only some provided insight along the way; after over 2 years, someone was able to pin down the issue and provide the solution. In the end, only a subset of the initial developers assigned to this bug were required to solve it.

In large organizations, there can be thousands of bugs generated every week with thousands of developers to pick from, each with different technical specialties. Some organizations can afford to hire human bug triagers. However, they may not have the background knowledge required to determine the correct developer; if they can, there are still many hours spent in bug assignment and, given a mis-assignment, many developer hours wasted assessing cases and waiting for responses \cite{improve_bug_triage,reducing_effort_triage}.

\section{Machine Learning Model} \label{data}
The dataset employed in this study, $D = \{(x^i, y^i), x^i \in \mathbb{X}, y^i \in \mathbb{Y}\}_{i=1}^m$, was composed of $m =$ 236,865 entries representing cases ranging from 2018-10-30 to 2019-03-19. Here, $\{x^i\}$ and $\{y^i\}$ refer to inputs and targets, respectively, drawn from their sample spaces and collected in $\mathbb{X}$ and $\mathbb{Y}$, respectively. 

The failure data for a bug, $x$ (no superscript implies a generic case), is similar to information provided in human-generated bug cases and includes the operating system, regression traceback, etc., where there were 55 features in total. These features were textual and numerical with 9,462 tokens before pre-processing, which previously grew in number by 5\% to 10\% each month. To reduce overfitting and training time, we applied latent semantic analysis (LSA \cite{lsa-paper}). In this way, we were able to reduce the input vector dimensionality to 1,000 before observing any significant losses in accuracy. LSA was chosen over other approaches (word2Vec \cite{word2vec}, Brown Clustering \cite{brown}, etc.) due to the semi-structured nature of the failure data, with key topics being stored in certain features (e.g. OS, error code, etc.).

The owner history of a bug is contained in its target, $y$. By applying owner-importance-weighted labeling (see Section~\ref{heuristic}), we transformed each target, $y^i \in \mathbb{Y}$, into two pmfs, one over all teams and one over all developers. Each team was represented by its manager.  An org chart was used to retrieve the manager for each developer (see Section~\ref{OrgChart}).

\subsection{Dual DNN} \label{DNN}
\begin{figure}[h!]
\centering
\includegraphics[scale=0.3]{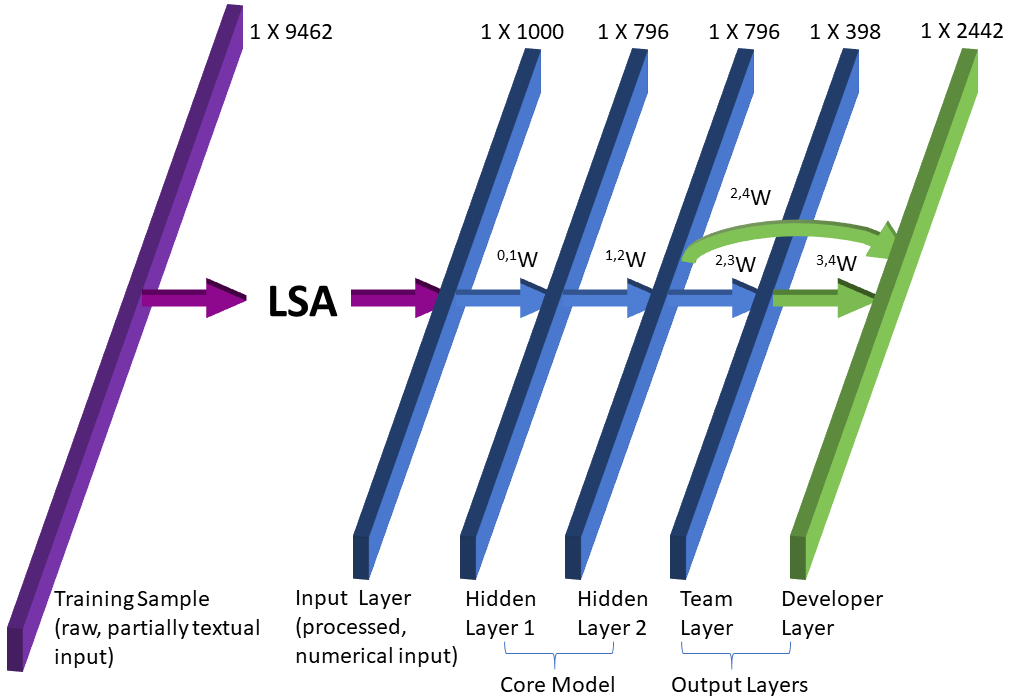} 
\caption{Dual DNN architecture with LSA preprocessing. Layer dimensions are shown at the top of each layer.}
\label{fig:neuralnet}
\end{figure}

The Dual DNN architecture employed for this work is shown in Figure~\ref{fig:neuralnet}. The weights are third-order tensors where $\prescript{k,l}{}{W}^{i,j}$ is the weight from layer $k$ to layer $l$, from neuron $i$ to neuron $j$. Each input vector was projected into a latent-space representation using LSA. The resulting latent vector (which is passed into the ``Input Layer'' in Figure~\ref{fig:neuralnet}) was then fed through two fully connected hidden layers (core model) where the number of neurons was parametrized to twice the number of teams. Holdout validation was used to optimize the hyperparameters of the Dual DNN. The Leaky Relu activation function defined as $y = x$ $\forall x \geq 0$ and $ y = 0.2x$ $\forall x < 0$ was used for each hidden layer. Dropout regularization with $p_\text{dropout}=0.1$ was used for each hidden layer. The final layer of the core model was then fed to each of the two output layers. The org chart-encoding scheme used to construct the output layers is explained in Section~\ref{OrgChart}.
The team layer constituted the first of the two outputs, providing an excellent fall-back for the model as teams are generally easier to predict than their developers (cf.~Section~\ref{results}). The team logits were fully connected to the second hidden layer and calculated as
\begin{equation}
    \text{team}\_\text{logit}_j = \sum_{h=0}^{n_\text{H}} \prescript{2,3}{}{W}^{h,j} \ ,
    \label{eqn:pred_team}
\end{equation}
where $n_\text{H}$ represents the number of neurons in the last hidden layer. The developer layer was fully connected to both the team layer and the second hidden layer, 
\begin{align}
    \text{developer}\_\text{logit}_j &= \sum_{h=0}^{n_\text{H}} \prescript{2,4}{}{W}^{h,j} + \sum_{t=0}^{n_\text{T}} \prescript{3,4}{}{W}^{t,j}%*1(person, person_j in team_t) \ 
    \label{eqn:pred_person} \ ,
\end{align}
where $n_\text{T}$ represents the number of teams.

\subsection{Org Chart Output Encoding} \label{OrgChart}
We designed Dual DNN to model an organization's flow of information by assuming that each team focuses on concrete products or topics within an organization that can be linked to a developer's technical specialty. The output encoding scheme was specifically designed to improve cost function optimization by using the model's own team prediction to aid its developer prediction. This novel scheme is constructed at model creation time based on a pre-computed graphical representation of the org chart. This process increases model robustness by enabling easier transfer learning, as explained in Section~\ref{graph}.

\subsubsection{Org Chart Design} \label{orgdesign}
The goal of the dual-output scheme was to allow the core model to learn which groups of bugs are owned by which team, which developer, and how these two correlate. The team output corresponds to a list of the managers in the org chart identified by level-order traversal. Here, each manager is assumed to represent a functional team within the organization. The developer output was created by traversing the manager array and, in order, placing their direct subordinates.

\subsubsection{Model Robustness} \label{graph}
As the model output was encoded using the org chart, any organizational changes can be directly reflected in the machine learning model. For instance, a developer being promoted to a new team will be captured in the org chart. This developer's neuron position can be updated, transferring its previous team--developer weights and accepting the new bug--developer weights. This architecture was designed to aid transfer learning in these scenarios, as team--developer correlations can be maintained despite role changes.

\subsubsection{Staged Learning} \label{staged}
With two different outputs stemming from the same hidden layers, training on the team and developer labels will each backpropogate gradients to the core model. First, the model was trained on the team outputs, which provided a less sparse output to learn on. Then, a second round of training was performed on the developer labels to learn bug--developer relationships and team--developer relationships. In the second round, the hidden layers were fixed so that their weights may still align with the team output layer.

\subsection{Owner-Importance-Weighted Labeling}\label{heuristic}

Each target in $\mathbb{Y}$ was numerically encoded by considering the contribution of each developer to resolving the associated bug. This contribution was quantified by assigning each of a developer's responses to one of three categories~--- the list of owners (excluding the closer), the list of commenters, and the closer --- and then counting the responses per developer in each category (cf.~Algorithm~\ref{alg:heuristic}, where `Y' and `m' refer to the set $\mathbb{Y}$ and the number of entries in $D$, respectively, and `n\_people' represents the number of either teams or individual developers).

\begin{algorithm}
    \caption{Encoding targets from categorized owner list.}
    \label{alg:heuristic}
    \begin{algorithmic}[1]
        \Procedure{make\_labels}{Y,
        n\_people}
            \For{i $\gets$ 0 to m}
                \State base $\gets$ zeros(n\_people)
                \For{$j \gets$ 0  to len(Y[i].owner)}
                    \State base[Y[i].owner[j]] += owner\_weight
                \EndFor
                \For{j $\gets$ 0  to len(Y[i].commenter)}
                    \State {\small base[Y[i].commenter[j]] += commenter\_weight}
                \EndFor
                \State base[Y[i].closer] += closer\_weight
                \State Y[i] $\gets$ base
            \EndFor
        \EndProcedure
    \end{algorithmic}
\end{algorithm}

In the traditional multi-class classification approach, one-hot-encoded targets, $y_\text{one-hot}$, would be generated subsequent to Algorithm~\ref{alg:heuristic}; instead, we transformed the targets by applying the softmax function,
  \begin{equation}
    \sigma{(o_j)} = \frac{e^{\frac{o_j}{T}}}{\sum_{k=1}^{n}e^{\frac{o_k}{T}}} \ , 
    \label{eqn:softmax} 
\end{equation}
where $o_j$ is traditionally the $j$th logit and $T$ is the temperature, a nonlinear scaling factor. To generate a softmax-transformed target, $y_\text{softmax}$, we used numerically encoded labels, \{$y_j$\}, in place of logits, \{$o_j$\}. Hence, $y_{\text{softmax},j} = \sigma(y_j)$. Here, $T$ was indirectly set by the weights in Algorithm~\ref{alg:heuristic} and separately for each class,
\begin{eqnarray}
    r_{\text{owner},j} &=& c_{\text{owner},j} \cdot \text{owner\_weight}\nonumber \ , \\
    r_{\text{commenter},j} &=& c_{\text{commenter},j} \cdot \text{commenter\_weight}\nonumber \ , \\
    r_{\text{closer},j} &=& c_{\text{closer},j} \cdot \text{closer\_weight}\nonumber \ , \\
    T_j &=& \dfrac{c_{\text{owner},j}+c_{\text{commenter},j}+c_{\text{closer},j}}{r_{\text{owner},j}+r_{\text{commenter},j}+r_{\text{closer},j}} \ ,
    \label{t}
\end{eqnarray}
where $r_{\text{category},j}$ is the contribution of the $j$th developer to a bug in a given category~$:=$~\{owner, commenter, closer\} and $c_{\text{category},j}$ is the number of that developer's responses to a bug in that category. Note that in our approach, the traditionally global temperature parameter $T$ introduced in Eq.~(\ref{eqn:softmax}) is an index-specific parameter.

Switching the traditional multi-class, one-hot targets, $y_\text{one-hot}$, with our softmax targets, $y_\text{softmax}$, directly impacts the gradient of the loss function. We employed the cross-entropy loss function,
\begin{equation}
    L = -\frac{1}{n}\sum_{j=1}^{n}y_{\text{distr},j}\log\big(\sigma(o_{j})\big) \ ,
    \label{eqn:crossentropyloss}
\end{equation}
where distr~$:=$~\{one-hot, softmax\}.
The general gradient of the loss along the $k$th dimension is
\begin{equation}
    \frac{\partial L}{\partial o_{k}} = -\frac{1}{n}\sum_{j=1}^{n}y_{\text{distr},j}\big(\delta_{jk}-\sigma{(o_k})\big)\label{eqn:normalgradient} \ ,
\end{equation}
where $\delta_{jk}$ is the Kronecker delta, i.e., $\delta_{jk}=1$ if $j=k$ and $\delta_{jk}=0$ otherwise.

In the multi-class case (distr~$=$~one-hot), this expression reduces to 
\begin{equation}
\label{eqn:multiclassgradient}
    \frac{\partial L}{\partial o_{k}} =
    \begin{cases}
    -\frac{y_{p}}{n}\big(1-\sigma{(o_{k}})\big), & k=p\\
    \frac{y_{p}}{n}\sigma{(o_{k}}), & k\neq p
    \end{cases} \ ,
\end{equation}
where $y_p$ represents the single positive label (i.e., $y_j = \delta_{jp}$).
Consequently, the gradient is only backpropagated with respect to that positive label, $y_p$.
The gradient along the $p$th dimension increases when the probability $\sigma(o_p)$ is close to zero, whereas the gradient along all other dimensions increases when $\sigma(o_{j\neq p})$ is close to one. This approach is not ideal as the model only learns from the single positive label, which is assumed to be the correct owner. 

In our multi-label case (distr~$=$~softmax), all labels \{$y_{\text{softmax},j}$\} take strictly non-zero values (probabilities), which relaxes the assumption that the positive label (here, the one associated with the highest probability) \textit{must} refer to the most suitable owner for a bug.
Instead, any developer may own a bug with a certain probability, where each label $y_{\text{softmax},j}$ in a target reflects that probability. Because ${y_{\text{softmax},j} > 0 \ \forall j}$, the full cross-entropy loss gradient, shown in Eq.~(\ref{eqn:normalgradient}), applies.  Hence, the gradient is backpropagated with respect to all $n$ classes, i.e., the gradient along any dimension is the probability-weighted sum of the dimension's own logit's activation with every other logit's activation. Contrary to the multi-class approach, the model is informed by the probabilistic contribution of each label instead of learning only from a single label.

In Section~\ref{results}, we show that the full cross-entropy loss gradient following from the multi-label classification approach yields an improved accuracy compared with previous studies employing the (single-label) multi-class classification approach. 

\section{Results} \label{results}
For a given input vector, $x$, our model outputs two pmf predictions, one over all teams and one over all developers. To provide a realistic use case where the model is required to assign future bugs based on prior knowledge, the dataset entries were time-sorted by bug submission date. To evaluate the top-$k$ assignment accuracy, we compared the top-$k$ predicted classes ($k$ highest probabilities) with the top-1 target labels,
\begin{equation}
    \text{Acc\%}_k = \frac{100}{m} \sum_{i=1}^{m} \left( {^{k}}y^{\ast,i}\right)^{\top} {^{1}}y^{i} = \frac{100}{m} \left( {^{k}}\mathbb{Y}^{\ast}\right)^{\top} {^{1}}\mathbb{Y} \ .
\end{equation}
Here, $^1y^i$ is a one-hot-encoded column vector of the multi-label target $y_\text{softmax}^i$, where the highest-probability label is set to one and every other label is set to zero. Similarly, $^ky^{\ast,i}$ is a $k$-hot-encoded column vector of the pmf prediction $y^{\ast,i}$, i.e., the top-$k$ predicted classes are set to one and every other class is set to zero. $^1\mathbb{Y}$ and $^k\mathbb{Y}^\ast$ are matrix generalizations of the vectors $^1y^i$ and $^ky^{\ast,i}$, respectively. There is at most a single element $j$ for which $^1y^i_j = {^ky^{\ast,i}_j} = 1$, which refers to a successful assignment. With an increasing $k$-value, the prediction/assignment accuracy will increase. However, since the metric does not require the model to discriminate between the suitability of the top-$k$ developers/teams, increasing $k$ decreases the usefulness of the metric in evaluating the model for a practical setting.

We compared the prediction performance of Dual DNN with that of random forests, logistic regression, and multinomial naive Bayes. We report results for holdout testing (note that there is an additional holdout validation set for hyperparameter tuning) for team prediction accuracy in Tables~\ref{tab:standardteam}~and~\ref{tab:standardteamnoheuristic}, and for developer prediction accuracy in Tables~\ref{tab:standardperson}~and~\ref{tab:standardpersonnoheuristic}. We as well report 11-fold IL-CV \cite{developerprioritization,deep_triage,tossinggraphs} results for team prediction accuracy in Tables~\ref{tab:kfoldteam}~and~\ref{tab:kfoldteamnoheuristic}, and for developer prediction accuracy in Tables~\ref{tab:kfoldperson}~and~\ref{tab:kfoldpersonnoheuristic}. The optimal owner importance weights were determined by holdout validation, with owner\_weight $=$ commenter\_weight $=$ closer\_weight $=0.5$, where all owner-importance-weighted results are based on these weights. Note that the temperature, $T$, increases with decreasing values of these weights, leading to a softmax function that is less concentrated around high-probability labels (cf.~Eq.~(\ref{eqn:softmax})). Though holdout validation was used to tune the hyperparameters, we rely on 11-fold IL-CV as a more accurate measure of the models' performances similar to previous studies \cite{ensemble,nb_svm_bug_tossing,deep_triage} (see Tables~\ref{tab:kfoldteam}~and~\ref{tab:kfoldperson}).

\subsection{Owner-Importance-Weighted Labeling} \label{res:ownerimportance}
To showcase the performance improvement using owner-importance-weighted labels, we conducted the learning and prediction tasks mentioned above twice: once with owner importance weighting and once without. The latter aligns with the labeling employed in previous studies \cite{ensemble,improve_bug_triage,who_should_fix_bug,nb_svm_bug_tossing,deep_triage,reducing_effort_triage}. Table~\ref{tab:heuristicimpact} showcases both the net performance improvements for all models and separately for Dual DNN when employing owner-importance-weighted labeling compared with the previous approaches. Dual DNN is separated as it is based on a multi-label learning scheme as opposed to the reference models which use a multi-class learning scheme. Note that the multi-class targets used here are different from previous studies as they were derived from the owner-importance-weighted labels; there was an observed 6\% deviation in the highest-probability labels. 

\begin{table}[!ht]
\caption{Increase of top-10 assignment accuracy, based on 11-fold IL-CV, when using owner-importance-weighted labels instead of labels that are not owner-importance-weighted. Random forests, logistic regression, multinomial naive Bayes, and Dual DNN were considered.}
    \begin{center}
     \begin{tabular}{|| c || c | c ||} 
     \hline
    Dataset & Average increase & Dual DNN increase \\ [0.5ex]
     \hline\hline 
     Team & 4\%-points & 11\%-points \\
     \hline
     Developer & 4\%-points & 16\%-points \\
     \hline
    \end{tabular}
    \end{center}
    \label{tab:heuristicimpact}
\end{table}

The performance accuracy increase was largest for the Dual DNN architecture, with 11\%-points for team assignments and 16\%-points for developer assignments. This finding provides evidence that the multi-label learning scheme exhibits synergistic effects with owner-importance-weighted labeling by enabling the model to learn from the full representation of the pmf target, $y_\text{softmax}$. Not only does the model-averaged performance increase, but every model's accuracy improves when employing owner-importance-weighted labels instead of their non-weighted analogs.

\subsection{Relative Performance of Dual DNN}\label{res:dualdnn}
We observed that the Dual DNN architecture outperformed all other models across all tests.
In developer assignment, the Dual DNN architecture outperformed all reference models by 21\%-points to 27\%-points. In addition, we compared the developer prediction accuracy of Dual DNN with that of a single-output DNN, denoted Developer DNN, which lacks the team output layer but is otherwise identical to Dual DNN (see Tables~\ref{tab:standardperson},\ref{tab:standardpersonnoheuristic},\ref{tab:kfoldperson},~and~\ref{tab:kfoldpersonnoheuristic}). Dual DNN outperformed Developer DNN by about 12\%-points, with the largest improvement in prediction accuracy occurring when using owner-importance-weighted labels. Further, Dual DNN consistently outperformed the reference models in team assignment with a performance increase of 10\%-points to 16\%-points.

\section{Conclusion and Outlook}
In this paper, we introduced an automated bug triager based on a novel two-output DNN architecture (Dual DNN) combined with LSA-projected inputs of failure data. Dual DNN exceeds all baselines in both team and developer triaging, utilizing its own knowledge of the team classes (team logit values) to aid in developer assignment. This scheme learns from a latent topic representation of 1,000 dimensions derived, utilizing LSA, from a corpus of 9,462 tokens. Since the output layers are encoded using the org chart, this model is robust against organizational changes because it is enabled to maintain team--developer relationships in the event of role changes. On average, the model is shown to beat baselines by 14\%-points and 25\%-points with respect to team and developer prediction, respectively. 

In addition to the Dual DNN architecture, we introduced and propose using a multi-label classification approach by weighting the number and type of a developer's responses to a bug to determine the owner importance in labeling. This approach was utilized to improve the targets used for this supervised learning problem and is a step toward aligning them with the actual contributions of a developer on a given case. To assess the performance of owner-importance-weighted labels, we considered additional reference models based on random forests, logistic regression, and multinomial naive Bayes. Compared with non-weighted labels, owner-importance-weighted labels led to a model-averaged improvement of 4\%-points in both team and developer assignment accuracy. Considering Dual DNN only, owner-importance-weighted labeling improved results by as much as 16\%-points. Note that our approach is different from the one employed by Xi et al. \cite{seqtriage}, which offers an improvement in label quality by taking into account the sequence of owners, but not their contribution. 

There are several possible improvements to our approach that will be explored in future work. In particular, other NLP methods (e.g., Latent Direchlet Allocation \cite{lda} and Transformer-XL \cite{transformerxl}), the heuristic employed for owner-importance-weighted labeling (e.g., including code repository commits), and other multi-label activation functions (e.g., sparsemax \cite{sparsemax}) could be explored. Moreover, we propose to extend this architecture's ties to the org chart to increase model robustness by directly altering the model in response to different organizational changes.

\section*{Acknowledgments}
J.P.~acknowledges funding through an \textit{Early Postdoc.Mobility} fellowship by the Swiss National Science Foundation (project~no.\,178463).

\begin{table*}[!ht]
\caption{Top-1,2,3,5,10 test set accuracies for team assignment based on holdout testing utilizing a 80:10:10 (training:validation:test) split with owner-importance-weighted labels, where owner\_weight $=$ commenter\_weight $=$ closer\_weight $=0.5$.}
    \begin{center}
     \begin{tabular}{|| c | c c c c c ||} 
     \hline
     Model & top-1 & top-2 & top-3 & top-5 & top-10 \\ [0.5ex]
     \hline\hline
     Dual DNN & $77.59 \pm 1.25$ & $80.64 \pm 0.65$ & $83.08 \pm 1.20$ & $86.90 \pm 0.53$ & $88.91 \pm 0.45$ \\ 
     \hline
     Logistic Regression & $3.06 \pml 0.00$ & $4.46 \pml 0.00$ & $7.29 \pml 0.00$ & $13.90 \pml 0.00$ & $21.98 \pml 0.00$ \\
     \hline
     Random Forest & $3.58 \pm 0.48$ & $12.97 \pm 2.41$ & $22.73 \pm 4.04$ & $32.01 \pm 2.33$ & $50.21 \pm 22.39$ \\
     \hline
     Multinomial Bayes & $2.94 \pml 0.00 $ & $4.39 \pml 0.00$ & $11.28 \pml 0.00$ & $32.06 \pml 0.00$ & $36.49 \pml 0.00$ \\
     \hline
    \end{tabular}
    \end{center}
    \label{tab:standardteam}
\end{table*}
\begin{table*}[!ht]
\caption{Top-1,2,3,5,10 test set accuracies for developer assignment based on holdout testing (80:10:10 split). Owner-importance-weighted labels were used (cf.~Table~\ref{tab:standardteam}).}
    \begin{center}
     \begin{tabular}{|| c | c c c c c ||} 
     \hline
     Model & top-1 & top-2 & top-3 & top-5 & top-10 \\ [0.5ex]
     \hline\hline
     Dual DNN & $4.51 \pm 0.37$ & $5.41 \pm 0.10$ & $5.75 \pm 0.21$ & $8.04 \pm 0.88$ & $14.08 \pm 1.05$ \\ 
     \hline
     Developer DNN & $3.70 \pm 0.36$ & $4.88 \pm 0.45$ & $5.65 \pm 0.17$ & $7.36 \pm 0.2$ & $8.77 \pm 0.05$ \\
     \hline
     Logistic Regression & $2.83 \pml 0.00$ & $3.67 \pml 0.00$ & $4.01 \pml 0.00$ & $4.45 \pml 0.00$ & $ 5.02 \pml 0.00$ \\
     \hline
     Random Forest & $2.71 \pm 0.26$ & $3.33 \pm 0.07$ & $3.60 \pm 0.14$ & $3.84 \pm 0.18$ & $5.03 \pm 0.92$ \\
     \hline
     Multinomial Bayes & $1.90 \pml 0.00$ & $2.53 \pml 0.00$ & $2.70 \pml 0.00$ & $4.04 \pml 0.00$ & $4.00 \pml 0.00$ \\
     \hline
    \end{tabular}
    \end{center}
    \label{tab:standardperson}
\end{table*}

\begingroup
\setlength{\tabcolsep}{5pt}
\begin{table*}[!ht]
\caption{Top-10 test set accuracies for team assignment based on 11-fold IL-CV. Owner-importance-weighted labels were used (cf.~Table~\ref{tab:standardteam}).}
    \begin{center}
     \begin{tabular*}{\textwidth}{|| c@{\extracolsep{\fill}} | c c c c c ||} 
     \hline
     Model & 1 & 2 & 3 & 4 & 5 \\ [0.5ex]
     \hline\hline
     Dual DNN & $77.92 \pm 2.01$ & $84.79 \pm 1.08$ & $74.13 \pm 1.78$ & $79.65 \pm 1.54$ & $72.56 \pm 1.02$ \\ 
     \hline
     Logistic Regression & $72.27 \pml 0.00$ & $80.27 \pml 0.00$ & $76.41 \pml 0.00$ & $69.41 \pml 0.00$ & $51.96 \pml 0.00$ \\
     \hline
     Random Forests & $72.56 \pm 0.93$ & $75.28 \pm 2.32$ & $75.23 \pm 1.68$ & $64.59 \pm 9.05$ & $61.78 \pm 10.69$ \\
     \hline
     Multinomial Bayes & $68.42 \pml 0.00$ & $68.05 \pml 0.00$ & $64.60 \pml 0.00$ & $60.10 \pml 0.00$ & $56.84 \pml 0.00$ \\
     \hline\hline
    \end{tabular*}
    \begin{tabular*}{\textwidth}{|| c@{\extracolsep{\fill}} | c c c c c >{\bfseries}c ||} 
     \hline
     Model & 6 & 7 & 8 & 9 & 10 & Average \\ [0.5ex]
     \hline\hline
     Dual DNN & $83.69 \pm 1.01$ & $83.08 \pm 0.58$ & $78.43 \pm 0.89$ & $87.61 \pm 0.54$ & $38.78 \pm 0.78$ & 76.07 \\ 
     \hline
     Logistic Regression & $50.12 \pml 0.00$ & $48.26 \pml 0.00$ & $33.72 \pml 0.00$ & $46.99 \pml 0.00$ & $70.84 \pml 0.00$ & 60.03 \\
     \hline
     Random Forests & $70.59 \pm 10.69$ & $53.13 \pm 1.85$ & $56.43 \pm 16.80$ & $42.01 \pm 18.44$ & $95.55 \pm 0.64$ & 66.72 \\
     \hline
     Multinomial Bayes & $59.22 \pml 0.00$ & $54.69 \pml 0.00$ & $47.08 \pml 0.00$ & $27.75 \pml 0.00$ & $93.38 \pml 0.00$ & 60.01 \\
     \hline
    \end{tabular*}
    \end{center}
    \label{tab:kfoldteam}
\end{table*}
\endgroup
\begingroup
\setlength{\tabcolsep}{4pt}
\begin{table*}[!ht]
\caption{Top-10 test set accuracies for developer assignment based on 11-fold IL-CV. Owner-importance-weighted labels were used (cf.~Table~\ref{tab:standardteam}).}
    \begin{center}
     \begin{tabular*}{\textwidth}{|| c@{\extracolsep{\fill}} | c c c c c ||} 
     \hline
     Model & 1 & 2 & 3 & 4 & 5 \\ [0.5ex]
     \hline\hline
     Dual DNN & $58.02 \pm 1.98$ & $59.63 \pm 1.67$ & $48.66 \pm 1.79$ & $54.68 \pm 1.34$ & $49.56 \pm 1.44$ \\
     \hline
     Developer DNN & $51.37 \pm 2.01$ & $64.87 \pm 1.54$ & $43.23 \pm 1.78$ & $26.16 \pm 2.05$ & $49.16 \pm 1.49$ \\
    
     \hline
     Logistic Regression & $43.32 \pml 0.00$ & $40.34 \pml 0.00$ & $39.44 \pml 0.00$ & $32.44 \pml 0.00$ & $30.47 \pml 0.00$ \\
     \hline
     
     Random Forests & $29.66 \pm 1.22$ & $27.67 \pm 0.75$ & $32.37 \pm 4.38$ & $30.37 \pm 0.87$ & $29.50 \pm 0.75$ \\
     \hline
     Multinomial Bayes & $24.02 \pml 0.00$ & $27.00 \pml 0.00$ & $30.06 \pml 0.00$ & $37.57 \pml 0.00$ & $31.01 \pml 0.00$ \\
     \hline\hline
    \end{tabular*}
    \begin{tabular*}{\textwidth}{|| c@{\extracolsep{\fill}} | c c c c c >{\bfseries}c ||} 
     \hline
     Model & 6 & 7 & 8 & 9 & 10 & Average \\ [0.5ex]
     \hline\hline
     Dual DNN & $69.37 \pm 1.21$ & $58.55 \pm 1.45$ & $67.4 \pm 1.12$ & $79.70 \pm 0.98$ & $6.86 \pm 0.87$ & 55.24 \\
     \hline
     Developer DNN & $64.65 \pm 1.54$ & $67.65 \pm 1.43$ & $49.81 \pm 1.23$ & $13.07 \pm 1.01$ & $1.16 \pm 0.98$ & 43.11 \\
     \hline
     Logistic Regression & $34.66 \pml 0.00$ & $31.74 \pml 0.00$ & $23.82 \pml 0.00$ & $7.05 \pml 0.00$ & $59.66 \pml 0.00$ & 28.44 \\
     \hline
     
     Random Forests & $32.93 \pm 0.33$ & $32.24 \pm 0.40$ & $25.64 \pm 0.23$ & $6.53 \pm 0.13$ & $59.49 \pm 0.74$ & 30.64\\
     \hline
     Multinomial Bayes & $34.11 \pml 0.00$ & $33.58 \pml 0.00$ & $26.35 \pml 0.00$ & $6.66 \pml 0.00$ & $59.15 \pml 0.00$ & 30.95\\
     \hline
    \end{tabular*}
    \end{center}
    \label{tab:kfoldperson}
\end{table*}
\endgroup

\newpage
\begin{table*}[!ht]
\caption{Top-1,2,3,5,10 test set accuracies for team assignment based on holdout testing (80:10:10 split), without owner-importance-weighted labels.}
    \begin{center}
     \begin{tabular}{|| c | c c c c c ||} 
     \hline
     Model & top-1 & top-2 & top-3 & top-5 & top-10 \\ [0.5ex]
     \hline\hline
     Dual DNN & $31.39 \pm 0.98$ & $33.31 \pm 1.00$ & $34.32 \pm 1.11$ & $36.1 \pm 0.67$ & $37.24 \pm 0.57$ \\ 
     \hline
     Logistic Regression & $4.45 \pml 0.00$ & $5.61 \pml 0.00$ & $6.51 \pml 0.00$ & $12.92 \pml 0.00$ & $18.74 \pml 0.00$ \\
     \hline
     Random Forest & $5.17 \pm 0.58$ & $9.62 \pm 3.32$ & $15.0 \pm 4.58$ & $30.01 \pm 8.86$ & $47.32 \pm 16.0$ \\
     \hline
     Multinomial Bayes & $4.22 \pml 0.00$ & $5.89 \pml 0.00$ & $13.3 \pml 0.00$ & $34.53 \pml 0.00$ & $37.9 \pml 0.00$ \\
     \hline
    \end{tabular}
    \end{center}
    \label{tab:standardteamnoheuristic}
\end{table*}
\begin{table*}[!ht]
\caption{Top-1,2,3,5,10 test set accuracies for team assignment based on holdout testing (80:10:10 split), without owner-importance-weighted labels.}
    \begin{center}
     \begin{tabular}{|| c | c c c c c ||} 
     \hline
     Model & top-1 & top-2 & top-3 & top-5 & top-10 \\ [0.5ex]
     \hline\hline
     Dual DNN & $0.96 \pm 0.02$ & $2.02 \pm 0.16$ & $2.44 \pm 0.26$ & $7.79 \pm 5.11$ & $18.09 \pm 8.39$  \\ 
     \hline
     Developer DNN & $2.36 \pm 0.17$ & $3.48 \pm 0.11$ & $3.74 \pm 0.13$ & $4.10 \pm 0.14$ & $4.56 \pm 0.22$ \\
     \hline
     Logistic Regression & $3.95 \pml 0.00$ & $4.39 \pml 0.00$ & $4.56 \pml 0.00$ & $4.89 \pml 0.00$ & $5.30 \pml 0.00$ \\
     \hline
     Random Forests & $4.08 \pm 0.01$ & $4.36 \pm 0.05$ & $4.57 \pm 0.09$ & $4.86 \pm 0.29$ & $5.80 \pm 0.73$ \\
     \hline
     Multinomial Bayes & $2.91 \pml 0.00$ & $3.44 \pml 0.00$ & $3.69 \pml 0.00$ & $4.03 \pml 0.00$ & $4.66 \pml 0.00$ \\
     \hline
    \end{tabular}
    \end{center}
    \label{tab:standardpersonnoheuristic}
\end{table*}

\begingroup
\setlength{\tabcolsep}{8pt}
\begin{table*}[!ht]
\caption{Top-10 test set accuracies for team assignment based on 11-fold IL-CV, without owner-importance-weighted labels.}
    \begin{center}
     \begin{tabular*}{\textwidth}{|| c@{\extracolsep{\fill}} | c c c c c ||} 
     \hline
     Model & 1 & 2 & 3 & 4 & 5 \\
     \hline\hline
     Dual DNN & $83.55 \pm 4.58$ & $82.98 \pm 1.54$ & $77.67 \pm 2.31$ & $75.73 \pm 2.05$ & $73.65 \pm 1.74$ \\ 
     \hline
     Logistic Regression & $72.36 \pml 0.00$ & $81.0 \pml 0.00$ & $76.26 \pml 0.00$ & $69.02 \pml 0.00$ & $51.83 \pml 0.00$ \\
     \hline
     Random Forests &  $61.92 \pm 4.75$& $64.19 \pm 8.26$& $59.71 \pm 6.74$& $56.51 \pm 9.84$& $61.14 \pm 9.27$ \\
     \hline
     Multinomial Bayes & $68.42 \pml 0.00$ & $68.59 \pml 0.00$ & $64.63 \pml 0.00$ & $60.11 \pml 0.00$ & $57.60 \pml 0.00$ \\
     \hline\hline
    \end{tabular*}
    {\small
    \begin{tabular*}{\textwidth}{|| c@{\extracolsep{\fill}} | c c c c c >{\bfseries}c ||} 
     \hline
     Model & 6 & 7 & 8 & 9 & 10 & Average\\ [0.5ex]
     \hline\hline
     Dual DNN & $53.96 \pm 1.62$ & $50.8 \pm 0.94$ & $32.57 \pm 0.18$ & $97.66 \pm 0.45$ & $39.79 \pm 0.40$ & 65.49\\ 
     \hline
     Logistic Regression & $49.98 \pml 0.00$ & $48.66 \pml 0.00$ & $35.4 \pml 0.00$ & $33.28 \pml 0.00$ & $71.17 \pml 0.00$ & 58.90 \\
     \hline
     Random Forests &  $56.20 \pm 9.11$& $54.23 \pm 10.03$& $49.62 \pm 14.18$& $45.06 \pm 25.84$& $90.43 \pm 3.68$ & 59.90\\
     \hline
     Multinomial Bayes & $59.85 \pml 0.00$ & $55.32 \pml 0.00$ & $48.14 \pml 0.00$ & $30.85 \pml 0.00$ & $95.23 \pml 0.00$ & 60.87 \\
     \hline
    \end{tabular*}}
    \end{center}
    \label{tab:kfoldteamnoheuristic}
\end{table*}
\endgroup
\begingroup
\setlength{\tabcolsep}{4pt}
\begin{table*}[!ht]
\caption{Top-10 test set accuracies for developer assignment based on 11-fold IL-CV, without owner-importance-weighted labels.}
    \begin{center}
     \begin{tabular*}{\textwidth}{|| c@{\extracolsep{\fill}} | c c c c c ||} 
     \hline
     Model & 1 & 2 & 3 & 4 & 5 \\ [0.5ex]
     \hline\hline
     Dual DNN & $52.93 \pm 6.03$ & $58.76 \pm 6.07$ & $52.47 \pm 3.55$ & $47.27 \pm 5.18$ & $49.92 \pm 4.54$  \\
     \hline
     Developer DNN & $52.58 \pm 1.57$ & $70.91 \pm 2.25$ & $55.62 \pm 1.28$ & $45.4 \pm 1.19$ & $48.63 \pm 1.10$ \\
     \hline
     Logistic Regression & $43.39 \pml 0.00$ & $40.21 \pml 0.00$ & $39.09 \pml 0.00$ & $32.13 \pml 0.00$ & $30.73 \pml 0.00$ \\
     \hline
     Random Forests & $24.52 \pm 1.77$ & $24.72 \pm 1.19$ & $27.05 \pm 2.80$ & $26.78 \pm 1.22$ & $27.43 \pm 1.52$\\
     \hline
     Multinomial Bayes & $24.03 \pml 0.00$ & $25.97 \pml 0.00$ & $29.21 \pml 0.00$ & $36.76 \pml 0.00$ & $30.39 \pml 0.00$ \\
     \hline\hline
    \end{tabular*}
    \begin{tabular*}{\textwidth}{|| c@{\extracolsep{\fill}} | c c c c c >{\bfseries}c ||} 
     \hline
     Model & 6 & 7 & 8 & 9 & 10 & Average\\ [0.5ex]
     \hline\hline
     Dual DNN & $39.82 \pm 0.61$ & $35.70 \pm 0.86$ & $23.66 \pm 1.05$ & $16.60 \pm 6.04$ & $11.14 \pm 7.50$ & 38.83 \\ 
     \hline
     Developer DNN & $38.88 \pm 0.28$ & $33.91 \pm 0.55$ & $2.33 \pm 0.50$ & $5.84 \pm 0.42$ & $4.95 \pm 1.21$ & 35.91\\
     \hline
     Logistic Regression & $35.33 \pml 0.00$ & $32.45 \pml 0.00$ & $24.71 \pml 0.00$ & $7.04 \pml 0.00$ & $59.6 \pml 0.00$ & 34.47 \\
     \hline
     Random Forests & $32.32 \pm 0.51$ & $30.99 \pm 0.67$ & $25.15 \pm 0.93$ & $6.71 \pm 0.09$ & $59.52 \pm 0.39$ & 28.52\\
     \hline
     Multinomial Bayes & $34.44 \pml 0.00$ & $33.95 \pml 0.00$ & $26.73 \pml 0.00$ & $6.47 \pml 0.00$ & $58.99 \pml 0.00$ & 30.69 \\
     \hline
    \end{tabular*}
    \end{center}
    \label{tab:kfoldpersonnoheuristic}
\end{table*}
\endgroup

\newpage\leavevmode\thispagestyle{empty}\newpage
\newpage\leavevmode\thispagestyle{empty}\newpage
\newpage\leavevmode\thispagestyle{empty}\newpage
\newpage\leavevmode\thispagestyle{empty}\newpage
\bibliographystyle{ieeetr}
\newpage
\bibliography{references}

\begin{thebibliography}{10}

\bibitem{information_extraction_methods}
R.~Shokripour, Z.~M. Kasirun, S.~Zamani, and J.~Anvik, ``Automatic bug
  assignment using information extraction methods,'' in {\em 2012 International
  Conference on Advanced Computer Science Applications and Technologies
  (ACSAT)}, pp.~144--149, Nov 2012.

\bibitem{reducing_effort_triage}
J.~Anvik and G.~C. Murphy, ``Reducing the effort of bug report triage:
  Recommenders for development-oriented decisions,'' {\em ACM Trans. Softw.
  Eng. Methodol.}, vol.~20, pp.~10:1--10:35, Aug. 2011.

\bibitem{lsiapplied}
S.~N. {Ahsan}, J.~{Ferzund}, and F.~{Wotawa}, ``Automatic software bug triage
  system (bts) based on latent semantic indexing and support vector machine,''
  in {\em 2009 Fourth International Conference on Software Engineering
  Advances}, pp.~216--221, Sep. 2009.

\bibitem{improve_bug_triage}
G.~Jeong, S.~Kim, and T.~Zimmermann, ``Improving bug triage with bug tossing
  graphs,'' in {\em Proceedings of the the 7th Joint Meeting of the European
  Software Engineering Conference and the ACM SIGSOFT Symposium on The
  Foundations of Software Engineering}, ESEC/FSE '09, (New York, NY, USA),
  pp.~111--120, ACM, 2009.

\bibitem{nb_svm_bug_tossing}
P.~Bhattacharya, I.~Neamtiu, and C.~R. Shelton, ``Automated, highly-accurate,
  bug assignment using machine learning and tossing graphs,'' {\em Journal of
  Systems and Software}, vol.~85, no.~10, pp.~2275 -- 2292, 2012.
\newblock Automated Software Evolution.

\bibitem{who_should_fix_bug}
J.~Anvik, L.~Hiew, and G.~C. Murphy, ``Who should fix this bug?,'' in {\em
  Proceedings of the 28th International Conference on Software Engineering},
  ICSE '06, (New York, NY, USA), pp.~361--370, ACM, 2006.

\bibitem{ensemble}
L.~Jonsson, M.~Borg, D.~Broman, K.~Sandahl, S.~Eldh, and P.~Runeson,
  ``Automated bug assignment: Ensemble-based machine learning in large scale
  industrial contexts,'' {\em Empirical Software Engineering}, vol.~21, 09
  2015.

\bibitem{tailored_1}
X.~Xia, D.~Lo, X.~Wang, and B.~Zhou, ``Accurate developer recommendation for
  bug resolution,'' in {\em 2013 20th Working Conference on Reverse Engineering
  (WCRE)}, pp.~72--81, Oct 2013.

\bibitem{tailored_2}
X.~Xie, W.~Zhang, Y.~Yang, and Q.~Wang, ``Dretom: Developer recommendation
  based on topic models for bug resolution,'' pp.~19--28, 09 2012.

\bibitem{deep_triage}
S.~Mani, A.~Sankaran, and R.~Aralikatte, ``Deeptriage: Exploring the
  effectiveness of deep learning for bug triaging,'' {\em CoRR},
  vol.~abs/1801.01275, 2018.

\bibitem{tossinggraphs}
P.~{Bhattacharya} and I.~{Neamtiu}, ``Fine-grained incremental learning and
  multi-feature tossing graphs to improve bug triaging,'' in {\em 2010 IEEE
  International Conference on Software Maintenance}, pp.~1--10, Sep. 2010.

\bibitem{developerprioritization}
J.~Xuan, H.~Jiang, Z.~Ren, and W.~Zou, ``Developer prioritization in bug
  repositories,'' in {\em 2012 34th International Conference on Software
  Engineering (ICSE)}, pp.~25--35, IEEE, 2012.

\bibitem{semisupervised}
J.~Xuan, H.~Jiang, Z.~Ren, J.~Yan, and Z.~Luo, ``Automatic bug triage using
  semi-supervised text classification,'' {\em arXiv preprint arXiv:1704.04769},
  2017.

\bibitem{multilabelknn}
X.~{Xia}, D.~{Lo}, X.~{Wang}, and B.~{Zhou}, ``Accurate developer
  recommendation for bug resolution,'' in {\em 2013 20th Working Conference on
  Reverse Engineering (WCRE)}, pp.~72--81, Oct 2013.

\bibitem{seqtriage}
S.~Xi, Y.~Yao, X.~Xiao, F.~Xu, and J.~Lu, ``An effective approach for routing
  the bug reports to the right fixers,'' in {\em Proceedings of the Tenth
  Asia-Pacific Symposium on Internetware}, Internetware '18, (New York, NY,
  USA), pp.~11:1--11:10, ACM, 2018.

\bibitem{lsa-paper}
T.~K. Landauer, P.~W. Foltz, and D.~Laham, ``An introduction to latent semantic
  analysis,'' {\em Discourse Processes}, vol.~25, no.~2-3, pp.~259--284, 1998.

\bibitem{word2vec}
T.~Mikolov, I.~Sutskever, K.~Chen, G.~S. Corrado, and J.~Dean, ``Distributed
  representations of words and phrases and their compositionality,'' in {\em
  Advances in neural information processing systems}, pp.~3111--3119, 2013.

\bibitem{brown}
P.~F. Brown, P.~V. Desouza, R.~L. Mercer, V.~J.~D. Pietra, and J.~C. Lai,
  ``Class-based n-gram models of natural language,'' {\em Computational
  linguistics}, vol.~18, no.~4, pp.~467--479, 1992.

\bibitem{lda}
D.~M. Blei, A.~Y. Ng, and M.~I. Jordan, ``Latent dirichlet allocation,'' {\em
  J. Mach. Learn. Res.}, vol.~3, pp.~993--1022, Mar. 2003.

\bibitem{transformerxl}
Z.~Dai, Z.~Yang, Y.~Yang, J.~Carbonell, Q.~V. Le, and R.~Salakhutdinov,
  ``Transformer-xl: Attentive language models beyond a fixed-length context,''
  2019.

\bibitem{sparsemax}
A.~F.~T. Martins and R.~F. Astudillo, ``From softmax to sparsemax: A sparse
  model of attention and multi-label classification,'' in {\em Proceedings of
  the 33rd International Conference on International Conference on Machine
  Learning - Volume 48}, ICML'16, pp.~1614--1623, JMLR.org, 2016.

\end{thebibliography}
\end{document}